\documentstyle[aps,tighten]{revtex}
\begin{document}
\title{Spin-Flip Transistor}
\author{Gerrit E. W. Bauer, Yuli V. Nazarov}
\address{Delft University of Technology, Department of Applied Physics and DIMES,\\
Lorentzweg 1, 2628 CJ Delft}
\author{Arne Brataas}
\address{Harvard University, Lyman Laboratory of Physics, Cambridge, MA 02138}
\date{\today }
\maketitle

\begin{abstract}
The recently developed semiclassical theory for magnetoelectronic circuits
is applied to a transistor-like device consisting of a normal metal island
and three magnetic terminals. The electric current between source and drain
can be controlled by the magnetization of a ``base'' reservoir up to
distances of the order of the spin-flip diffusion length.
\end{abstract}

\section{Introduction}

Magnetoelectronics is concerned with the integration of ferromagnetic metals
into conventional electronic circuits which can be applied to {\it e.g.}
improved magnetic field sensors and non-volatile magnetic random access
memories (MRAMs). Different magnetoelectronic devices with three or four
terminals (``spin transistors'') have been realized \cite{Johnson,Jedema00}
and proposed \cite{Datta,Brataas00a}. In the present manuscript we will
analyze in detail the active magnetoelectronic device of Brataas {\it et al. 
}\cite{Brataas00a}, which we call the {\em spin-flip transistor}.

Transport in hybrid metallic systems in the presence of long-range
correlations in an order parameter can be described by a generalization of
Kirchhoff's theory of electronic circuits when the electronic phase is
sufficiently scrambled in parts of the system, the ``nodes''. This approach
has been pioneered in Ref. \cite{Nazarov} for electronic networks with
superconducting elements. It has recently been adopted also for
magnetoelectronic circuits \cite{Brataas00a}, like the Johnson spin
transistor \cite{Johnson} or the 4-terminal mesoscopic spin valve of Jedema 
{\it et al.} \cite{Jedema00}. The circuit theory can be derived from a given
Stoner Hamiltonian in terms of the Keldysh non-equilibrium Green function
formalism in spin space \cite{Brataas00b}. The basic physics is provided by
dividing the system into reservoirs, resistors and nodes which can be real
or fictitious. In order to arrive at a useful formalism, an isotropy
assumption has to be introduced for the nodes in which the electron
distributions may taken to be isotropic. This implies the presence of
sufficient disorder (or chaotic scattering). Inelastic\ or dephasing
scattering is not required but, at least in the nodes, not forbidden either.
Electron charges and spins are accumulated or depleted in the nodes as a
function of the applied voltages. Because the spin-accumulation is not
necessarily collinear to the spin-quantization axis, at each node the
electron distribution can be denoted as $\hat{f},$ where the hat ($\hat{}$)
denotes a $2\times 2$ matrix in spin-space. The external reservoirs are
assumed to be in local equilibrium so that the distribution matrix $\hat{f}=%
\hat{1}f_{\alpha }$ is diagonal in spin-space and equal to the local
chemical potential $f_{\alpha }$ in reservoir $\alpha $, where $\hat{1}$ is
the unit matrix. The direction of the magnetization of the ferromagnetic
nodes will be denoted by the unit vector ${\bf m}_{\alpha }$. The current
matrix $\hat{I}_{\alpha \beta }$ through a {\em contact} connecting two
neighboring nodes can be calculated as a function of the distribution
matrices on the adjacent nodes and the $2\times 2$ conductance tensor
composed of the spin-dependent conductances $G^{\uparrow }$ and $%
G^{\downarrow }$ 
\[
G^{s}=\frac{e^{2}}{h}\left[ M-\sum_{nm}|r_{s}^{nm}|^{2}\right] =\frac{e^{2}}{%
h}\sum_{nm}|t_{s}^{nm}|^{2}\,,
\]
and the {\em mixing conductance} 
\begin{equation}
G^{s,-s}=\frac{e^{2}}{h}\left[ M-\sum_{nm}r_{s}^{nm}(r_{-s}^{nm})^{\ast }%
\right] ,  \label{Gmixing}
\end{equation}
where $r_{s}^{nm},$ $t_{s}^{nm}$ are the reflection and transmission
coefficients in a spin-diagonal reference frame and $M$ the number of modes
in the absence of reflections. Spin-flips in the contacts have been
disregarded. In the following the node is taken to be a normal metal. When
its size in the transport direction is smaller than the spin-flip diffusion
length $l_{sf}=\sqrt{D\tau _{sf}}\ $(1$\mu $m in Cu \cite{Jedema00}), where $%
D$ is the diffusion coefficient, the spin-current conservation law 
\begin{equation}
\sum_{\alpha }\hat{I}_{\alpha \beta }=\left( \frac{\partial \hat{f}_{\beta
}^{N}}{\partial t}\right) _{\mbox{rel}}\,=\frac{\hat{1}\mbox{Tr}\hat{f}^{N}-2%
\hat{f}^{N}}{2\tau _{sf}}  \label{curcons}
\end{equation}
allows computation of the circuit properties as a function of the applied
voltages. The right hand side of Eq.~(\ref{curcons}) can be set to zero when
the spin-current in the node is conserved, {\em i.e.} when an electron
resides on the node sufficiently shorter than the spin-flip relaxation time $%
\tau _{sf}$.

In the following we concentrate on the practical application of the circuit
theory to the spin-flip transistor. We consider the limit of long $\tau
_{sf},$ which is experimentally relevant even at room temperature \cite
{Jedema00}. In order to simplify results we will confine attention to
half-metallic (fully polarized) ferromagnetic metals (HMF) where
appropriate. Our model for the interface between a normal metal and
half-metallic ferromagnet reads: 
\begin{equation}
\hat{G}=\left( 
\begin{array}{cc}
G & G \\ 
G & 0
\end{array}
\right) ,
\end{equation}
so $G^{\uparrow }=G^{\uparrow \downarrow }=G,\;G^{\downarrow }=0.$

\section{ Current matrix for F%
%TCIMACRO{\TEXTsymbol{\vert}}%
%BeginExpansion
\mbox{$\vert$}%
%EndExpansion
N contact}

The particle current on the normal side of an F%
%TCIMACRO{\TEXTsymbol{\vert}}%
%BeginExpansion
\mbox{$\vert$}%
%EndExpansion
N contact and directed into the normal metal is \cite{Brataas00a} 
\begin{equation}
\hat{I}^{\left( \vec{m}\right) }=G^{\uparrow }\hat{u}_{\vec{m}}^{\uparrow
}\left( \hat{f}^{F}-\hat{f}^{N}\right) \hat{u}_{\vec{m}}^{\uparrow
}+G^{\downarrow }\hat{u}_{\vec{m}}^{\downarrow }\left( \hat{f}^{F}-\hat{f}%
^{N}\right) \hat{u}_{\vec{m}}^{\downarrow }-G^{\uparrow \downarrow }\hat{u}_{%
\vec{m}}^{\uparrow }\hat{f}^{N}\hat{u}_{\vec{m}}^{\downarrow }-\left(
G^{\uparrow \downarrow }\right) ^{\ast }\hat{u}_{\vec{m}}^{\downarrow }\hat{f%
}^{N}\hat{u}_{\vec{m}}^{\uparrow },
\end{equation}
where 
\begin{equation}
\hat{u}_{\vec{m}}^{\uparrow }=\frac{1}{2}\left( \hat{1}+{\bf \hat{\sigma}}%
\cdot \vec{m}\right) ;\;\hat{u}_{\vec{m}}^{\downarrow }=\frac{1}{2}\left( 
\hat{1}-{\bf \hat{\sigma}}\cdot \vec{m}\right)
\end{equation}
are the spin-%
%TCIMACRO{\UNICODE{0xbd} }%
%BeginExpansion
$\frac12$%
%EndExpansion
\ rotation matrices. In the following we take the ferromagnet as a reservoir
at equilibrium $\hat{f}^{F}=f^{F}\hat{1}.$ Some insight can be gained by
re-writing the current and the distribution function in the form of a scalar
particle and a vectorial spin contribution. The distribution function on the
normal node can be written as 
\begin{equation}
\hat{f}^{N}=\left( 
\begin{array}{cc}
f^{\uparrow \uparrow } & f^{\uparrow \downarrow } \\ 
\left( f^{\uparrow \downarrow }\right) ^{\ast } & f^{\downarrow \downarrow }
\end{array}
\right) =C\hat{1}+{\bf S}\cdot \text{{\bf $\hat{\sigma}$}},
\end{equation}
where $C=\left( f^{\uparrow \uparrow }+f^{\downarrow \downarrow }\right) /2$
is the local chemical potential and \ 
\begin{equation}
{\bf S=}\frac{1}{2}\left( 
\begin{array}{c}
f^{\uparrow \downarrow }+\left( f^{\uparrow \downarrow }\right) ^{\ast } \\ 
f^{\uparrow \downarrow }-\left( f^{\uparrow \downarrow }\right) ^{\ast } \\ 
f^{\uparrow \uparrow }-f^{\downarrow \downarrow }
\end{array}
\right) =\left( 
\begin{array}{c}
%TCIMACRO{\func{Re}}%
%BeginExpansion
\mathop{\rm Re}%
%EndExpansion
f^{\uparrow \downarrow } \\ 
i%
%TCIMACRO{\func{Im}}%
%BeginExpansion
\mathop{\rm Im}%
%EndExpansion
f^{\uparrow \downarrow } \\ 
\frac{1}{2}\left( f^{\uparrow \uparrow }-f^{\downarrow \downarrow }\right)
\end{array}
\right) \equiv \left( 
\begin{array}{c}
M_{R} \\ 
M_{I} \\ 
\Sigma
\end{array}
\right)
\end{equation}
the spin accumulation vector. The junction parameters can be rewritten as $%
P=\left( G^{\uparrow }-G^{\downarrow }\right) /2,$ $G=G^{\uparrow
}+G^{\downarrow },$ $\Re =%
%TCIMACRO{\func{Re}}%
%BeginExpansion
\mathop{\rm Re}%
%EndExpansion
G^{\uparrow \downarrow },$ and $\Im =i%
%TCIMACRO{\func{Im}}%
%BeginExpansion
\mathop{\rm Im}%
%EndExpansion
G^{\uparrow \downarrow }.$ The matrix-current through an F%
%TCIMACRO{\TEXTsymbol{\vert}}%
%BeginExpansion
\mbox{$\vert$}%
%EndExpansion
N interface $\hat{I}=(I_{C}\hat{1}+{\bf I}_{s}\cdot {\bf \hat{\sigma}})/2$
can then be expanded into vector components in terms the scalar charge
current: 
\[
I_{C}=G\left( f^{F}-C\right) -2P{\bf m}\cdot {\bf S} 
\]
and the vector spin current: 
\begin{equation}
\frac{{\bf I}_{s}}{2}=[P(f^{F}-C)-\left( \frac{G}{2}-\Re \right) {\bf S}%
\cdot {\bf m}]{\bf m}-\Re {\bf S}-i\Im ({\bf S}\times {\bf m})
\end{equation}
The vector spin current component perpendicular to the magnetization
direction $-{\bf I}_{t}$ equals the spin-torque exerted by the polarized
current on the ferromagnet \cite{Katine00,Waintal} 
\begin{equation}
-{\bf I}_{\perp }=-{\bf I}_{s}+\left( {\bf I}_{s}\cdot {\bf m}\right) {\bf %
m=-}2%
%TCIMACRO{\func{Re}}%
%BeginExpansion
\mathop{\rm Re}%
%EndExpansion
G^{\uparrow \downarrow }{\bf \left( S\cdot m\right) m}+2\text{Re}G^{\uparrow
\downarrow }{\bf S}-2\text{Im}G^{\uparrow \downarrow }({\bf S}\times {\bf m}%
).
\end{equation}

When the magnetization is parallel to the quantization axis, $\vec{m}_{\pm
z}=(0,0,\pm 1):$%
\begin{equation}
\hat{u}_{z}^{\uparrow }=\left( \hat{1}+\hat{\sigma}_{z}\right) /2=\left( 
\begin{array}{cc}
1 & 0 \\ 
0 & 0
\end{array}
\right) ;\;\hat{u}_{z}^{\downarrow }=\left( \hat{1}-\hat{\sigma}_{z}\right)
/2=\left( 
\begin{array}{cc}
0 & 0 \\ 
0 & 1
\end{array}
\right)
\end{equation}
and $\hat{u}_{-z}^{\uparrow }=\hat{u}_{z}^{\downarrow }.$ It is easy to see
that 
\begin{equation}
\hat{I}^{\left( z\right) }=\left( 
\begin{array}{cc}
G^{\uparrow }\left( f^{F}-f_{\uparrow \uparrow }\right) & -G^{\uparrow
\downarrow }f_{\uparrow \downarrow } \\ 
-\left( G^{\uparrow \downarrow }f_{\uparrow \downarrow }\right) ^{\ast } & 
G^{\downarrow }\left( f^{F}-f_{\downarrow \downarrow }\right)
\end{array}
\right) ;\;\hat{I}^{\left( -z\right) }=\left( 
\begin{array}{cc}
G^{\uparrow }\left( f^{F}-f_{\downarrow \downarrow }\right) & -G^{\uparrow
\downarrow }\left( f_{\uparrow \downarrow }\right) ^{\ast } \\ 
-\left( G^{\uparrow \downarrow }\right) ^{\ast }f_{\uparrow \downarrow } & 
G^{\downarrow }\left( f^{F}-f_{\uparrow \uparrow }\right)
\end{array}
\right)
\end{equation}

Non-diagonal elements become important when the magnetization is not
collinear to the quantization axis. For $\vec{m}_{x}=(1,0,0)$ : 
\begin{equation}
\hat{u}_{x}^{\uparrow }=\left( \hat{1}+\hat{\sigma}_{x}\right) /2=\frac{1}{2}%
\left( 
\begin{array}{cc}
1 & 1 \\ 
1 & 1
\end{array}
\right) ;\;\hat{u}_{x}^{\downarrow }=\left( \hat{1}-\hat{\sigma}_{x}\right)
/2=\frac{1}{2}\left( 
\begin{array}{cc}
1 & -1 \\ 
-1 & 1
\end{array}
\right)
\end{equation}
and 
\begin{equation}
\hat{I}^{\left( x\right) }=\left( 
\begin{array}{cc}
\frac{G}{2}\left( f^{F}-C\right) -PM_{R}-\Re \Sigma +\Im M_{I} & P\left(
f^{F}-C\right) -\frac{G}{2}M_{R}+\Im \Sigma -\Re M_{I} \\ 
P\left( f^{F}-C\right) -\frac{G}{2}M_{R}-\Im \Sigma +\Re M_{I} & \frac{G}{2}%
\left( f^{F}-C\right) -PM_{R}+\Re \Sigma -\Im M_{I}
\end{array}
\right)  \label{xcurrent}
\end{equation}

\section{Two-terminal (spin valve) device}

In this section we determine the current through a  F%
%TCIMACRO{\TEXTsymbol{\vert}}%
%BeginExpansion
\mbox{$\vert$}%
%EndExpansion
N%
%TCIMACRO{\TEXTsymbol{\vert}}%
%BeginExpansion
\mbox{$\vert$}%
%EndExpansion
F structure with parallel, antiparallel and perpendicularly oriented
magnetizations, making use of the current conservation condition Eq. (\ref
{curcons}).  $f_{1}^{F}=\Delta \mu $ and $f_{2}^{F}=0.$

\subsection{Collinear magnetization}

Let us first determine $\hat{f}^{N}$ for parallel magnetizations in the
parallel F$_{1}^{\left( z\right) }$%
%TCIMACRO{\TEXTsymbol{\vert}}%
%BeginExpansion
\mbox{$\vert$}%
%EndExpansion
N%
%TCIMACRO{\TEXTsymbol{\vert}}%
%BeginExpansion
\mbox{$\vert$}%
%EndExpansion
F$_{2}^{\left( z\right) }$ configuration. Spin current conservation requires
that $\hat{I}_{1}^{\left( z\right) }+\hat{I}_{2}^{\left( z\right) }=0.$ 
\begin{equation}
\left( 
\begin{array}{cc}
G^{\uparrow }\left( \Delta \mu -f_{\uparrow \uparrow }\right)  & 
-G^{\uparrow \downarrow }f_{\uparrow \downarrow } \\ 
-\left( G^{\uparrow \downarrow }f_{\uparrow \downarrow }\right) ^{\ast } & 
G^{\downarrow }\left( \Delta \mu -f_{\downarrow \downarrow }\right) 
\end{array}
\right) =\left( 
\begin{array}{cc}
G^{\uparrow }f_{\uparrow \uparrow } & G^{\uparrow \downarrow }f_{\uparrow
\downarrow } \\ 
\left( G^{\uparrow \downarrow }f_{\uparrow \downarrow }\right) ^{\ast } & 
G^{\downarrow }f_{\downarrow \downarrow }
\end{array}
\right) 
\end{equation}
from which we conclude that $f_{\uparrow \downarrow }=0$ and $f_{\uparrow
\uparrow }=f_{\downarrow \downarrow }=\Delta \mu /2.$ 

When antiparallel F$_{1}^{\left( z\right) }$%
%TCIMACRO{\TEXTsymbol{\vert}}%
%BeginExpansion
\mbox{$\vert$}%
%EndExpansion
N%
%TCIMACRO{\TEXTsymbol{\vert}}%
%BeginExpansion
\mbox{$\vert$}%
%EndExpansion
F$_{2}^{\left( -z\right) },\;\hat{I}_{1}^{\left( z\right) }+\hat{I}%
_{2}^{\left( -z\right) }=0.$: 
\begin{equation}
\left( 
\begin{array}{cc}
G^{\uparrow }\left( \Delta \mu -f_{\uparrow \uparrow }\right)  & 
-G^{\uparrow \downarrow }f_{\uparrow \downarrow } \\ 
-\left( G^{\uparrow \downarrow }f_{\uparrow \downarrow }\right) ^{\ast } & 
G^{\downarrow }\left( \Delta \mu -f_{\downarrow \downarrow }\right) 
\end{array}
\right) =\left( 
\begin{array}{cc}
G^{\downarrow }f_{\uparrow \uparrow } & G^{\uparrow \downarrow }\left(
f_{\uparrow \downarrow }\right) ^{\ast } \\ 
\left( G^{\uparrow \downarrow }\right) ^{\ast }f_{\uparrow \downarrow } & 
G^{\uparrow }f_{\downarrow \downarrow }
\end{array}
\right) 
\end{equation}
\begin{equation}
f_{\uparrow \uparrow }=\frac{G^{\uparrow }\Delta \mu }{G^{\uparrow
}+G^{\downarrow }};\;f_{\downarrow \downarrow }=\frac{G^{\downarrow }\Delta
\mu }{G^{\uparrow }+G^{\downarrow }};\;\Sigma =\frac{P\Delta \mu }{G}
\end{equation}
Since $-f_{\uparrow \downarrow }=\left( f_{\uparrow \downarrow }\right)
^{\ast }\;%
%TCIMACRO{\func{Re}}%
%BeginExpansion
\mathop{\rm Re}%
%EndExpansion
f_{\uparrow \downarrow }=0$ and $%
%TCIMACRO{\func{Im}}%
%BeginExpansion
\mathop{\rm Im}%
%EndExpansion
f_{\uparrow \downarrow }$ is undetermined but irrelevant. The charge current 
\begin{equation}
I_{C}=G\Delta \mu -G^{\uparrow }f_{\uparrow \uparrow }-G^{\downarrow
}f_{\downarrow \downarrow }=\frac{2G^{\uparrow }G^{\downarrow }}{G^{\uparrow
}+G^{\downarrow }}\Delta \mu 
\end{equation}
vanishes for the HMF.

\subsection{Non-collinear}

$\hat{f}^{N}$ for normal magnetizations in F$_{1}^{\left( z\right) }$%
%TCIMACRO{\TEXTsymbol{\vert}}%
%BeginExpansion
\mbox{$\vert$}%
%EndExpansion
N%
%TCIMACRO{\TEXTsymbol{\vert}}%
%BeginExpansion
\mbox{$\vert$}%
%EndExpansion
F$_{2}^{\left( x\right) }$ structures follows again from the current
conservation condition $\hat{I}_{1}^{\left( z\right) }+\hat{I}_{2}^{\left(
x\right) }=0$ and using Eq. (\ref{xcurrent}): 
\begin{equation}
\left( 
\begin{array}{cc}
G^{\uparrow }\left( \Delta \mu -f_{\uparrow \uparrow }\right)  & 
-G^{\uparrow \downarrow }f_{\uparrow \downarrow } \\ 
-\left( G^{\uparrow \downarrow }f_{\uparrow \downarrow }\right) ^{\ast } & 
G^{\downarrow }\left( \Delta \mu -f_{\downarrow \downarrow }\right) 
\end{array}
\right) +\left( 
\begin{array}{cc}
-\frac{G}{2}C-PM_{R}-\Re \Sigma +\Im M_{I} & -PC-\frac{G}{2}M_{R}+\Im \Sigma
-\Re M_{I} \\ 
-PC-\frac{G}{2}M_{R}-\Im \Sigma +\Re M_{I} & -\frac{G}{2}C-PM_{R}+\Re \Sigma
-\Im M_{I}
\end{array}
\right) =0
\end{equation}
We find a spin accumulation vector 
\begin{equation}
{\bf S=}\frac{P\Delta \mu }{2}\frac{1}{\Re ^{2}-\Im ^{2}+\frac{G}{2}\Re }%
\left( 
\begin{array}{c}
-\Re  \\ 
\Im  \\ 
\Re 
\end{array}
\right) \stackrel{\text{HMF}}{\rightarrow }\left( 
\begin{array}{c}
-1 \\ 
0 \\ 
1
\end{array}
\right) \frac{\Delta \mu }{6}
\end{equation}
and a charge current (from left to right): 
\begin{equation}
I_{C}=G\left( \Delta \mu -C-\Sigma \right) =\frac{G}{2}-\frac{P^{2}}{\frac{G%
}{2}+\frac{\Re ^{2}-\Im ^{2}}{\Re }}\stackrel{\text{HMF}}{\rightarrow }G%
\frac{\Delta \mu }{3}.
\end{equation}

\section{ Three terminal device}

We now compute the characteristics of the device with three ferromagnetic
terminals attached to a normal metal node. The magnetization of the third
(base) terminal can be either collinear or normal to the magnetizations of
source and drain. The latter are invariably taken to be antiparallel.

\subsection{ Collinear configuration}

Let us consider a bias $\Delta \mu _{3}$ at which there is zero charge
current through the base terminal, {\it i.e.} $\left[ I_{\uparrow \uparrow
}+I_{\downarrow \downarrow }\right] _{3}=0$. For $\vec{m}_{3}=(0,0,1)$ this
translates into 
\begin{eqnarray}
0 &=&G^{\uparrow }\left( \Delta \mu _{3}^{\left( z\right) }-f_{\uparrow
\uparrow }\right) +G^{\downarrow }\left( \Delta \mu _{3}^{\left( z\right)
}-f_{\downarrow \downarrow }\right) \\
\Delta \mu _{3}^{\left( z\right) } &=&\frac{G^{\uparrow }f_{\uparrow
\uparrow }+G^{\downarrow }f_{\downarrow \downarrow }}{G^{\downarrow
}+G^{\downarrow }}=\frac{G^{\uparrow }\left( C+\Sigma \right) +G^{\downarrow
}\left( C-\Sigma \right) }{G^{\downarrow }+G^{\downarrow }}=C+\frac{2P\Sigma 
}{G}.
\end{eqnarray}
So in this case matrix-current conservation in the node reads: 
\begin{equation}
\hat{I}_{1}^{\left( z\right) }+\hat{I}_{2}^{\left( -z\right) }+\hat{I}%
_{3}^{\left( z\right) }=0
\end{equation}
We find: 
\begin{equation}
\Delta \mu _{3}^{\left( z\right) }=\frac{\left( G^{\downarrow }\right)
^{2}+G^{\downarrow }G^{\uparrow }+\left( G^{\uparrow }\right) ^{2}}{\left(
G^{\downarrow }\right) ^{2}+4G^{\downarrow }G^{\uparrow }+\left( G^{\uparrow
}\right) ^{2}}\Delta \mu
\end{equation}
The source-drain current 
\[
\frac{I_{C}^{\left( z\right) }}{G\Delta \mu }=\frac{3G^{\downarrow
}G^{\uparrow }}{\left( G^{\downarrow }\right) ^{2}+4G^{\downarrow
}G^{\uparrow }+\left( G^{\uparrow }\right) ^{2}} 
\]
vanishes again for HMF terminals$:$%
\begin{equation}
f_{\downarrow \downarrow }=0;\;f_{\uparrow \uparrow }=\Delta \mu
_{3}^{\left( z\right) }=\Delta \mu ;\;I_{C}^{\left( z\right) }=0;\;\Sigma =%
\frac{\Delta \mu }{2}.
\end{equation}

\subsection{Non-collinear configuration}

When the third electrode is rotated to the $x$-direction, the zero particle
current condition for the base contact dictates: 
\begin{equation}
\Delta \mu _{3}^{\left( x\right) }=C+2PM_{R}/G
\end{equation}
Matrix-current conservation: 
\begin{equation}
\hat{I}_{z}^{\left( 1\right) }+\hat{I}_{-z}^{\left( 2\right) }+\hat{I}%
_{x}^{\left( 3\right) }=0
\end{equation}
leads to: 
\begin{equation}
\Delta \mu _{3}^{\left( x\right) }=C=\Delta \mu /2
\end{equation}
\begin{equation}
I_{C}^{\left( x\right) }=\left( \frac{G}{2}-\frac{2P^{2}}{G+\left|
G^{\uparrow \downarrow }\right| ^{2}/%
%TCIMACRO{\func{Re}}%
%BeginExpansion
\mathop{\rm Re}%
%EndExpansion
G^{\uparrow \downarrow }}\right) \Delta \mu
\end{equation}
For the HMF, $\Sigma =\frac{\Delta \mu }{4}$ and $I_{C}^{\left( x\right) }=%
\frac{G}{4}\Delta \mu $. When (in this limit) the potential at the third
electrode is varied $\Sigma \ $does not change, but the length of ${\bf S}$
does: 
\begin{equation}
\left| {\bf S}\right| ^{2}=\Sigma ^{2}+M_{R}^{2}+M_{I}^{2}=\left( \frac{%
\Delta \mu }{4}\right) ^{2}+\left( \frac{2\Delta \mu ^{\left( 3\right)
}-\Delta \mu }{14}\right) ^{2}
\end{equation}
The spin-accumulation is minimal for the zero charge current condition for
the third terminal, which corresponds to a maximum of the spin-current
through the third terminal. The minimum spin-accumulation consequently
allows a maximum source-drain current without dissipation in the base.

\section{Discussion}

In the HMF limit the physics of the transistor action is easily understood.
In the collinear configuration the device is electrically dead: No current
can flow into the drain because spin-up states can not penetrate a HMF with
spin-down magnetization and source and base reservoir are at the same
potential. When the magnetization of the base is rotated by 90 degrees and
the potential is lowered to $\Delta \mu /2$ the incoming spin-up current is
exactly equal to the outflowing spin-down current. Although there is no
direct current between source and drain, the outflowing spin-down current
from the base can enter the drain. Effectively we thus have switched on a
source-drain charge current by rotating the magnetization. Since the base
contact operates as a perfect spin-flip, we suggest the name {\em spin-flip
transistor}{\it \ }for our device{\it .}

Let us consider the ideal case in which\ magnet number 3 has a negligible
anisotropy, thus can be rotated without appreciable\ energy cost. We still
need a voltage source for the base in order to stick to the optimal working
point. Let us consider the source-drain conductances in the three terminal
device for the $x$ (on) and $z$ (off) configurations: 
\begin{equation}
\frac{I_{C}^{\left( z\right) }}{\Delta \mu }=\frac{3G^{\downarrow
}G^{\uparrow }}{\left( G^{\downarrow }\right) ^{2}+4G^{\downarrow
}G^{\uparrow }+\left( G^{\uparrow }\right) ^{2}}G=6\frac{G^{2}-4P^{2}}{%
3G^{2}-4P^{2}}G\stackrel{HMF}{\rightarrow }0
\end{equation}
with 
\begin{equation}
\Delta \mu _{3}^{\left( z\right) }=\frac{\left( G^{\downarrow }\right)
^{2}+G^{\downarrow }G^{\uparrow }+\left( G^{\uparrow }\right) ^{2}}{\left(
G^{\downarrow }\right) ^{2}+4G^{\downarrow }G^{\uparrow }+\left( G^{\uparrow
}\right) ^{2}}\Delta \mu \stackrel{HMF}{\rightarrow }\Delta \mu 
\end{equation}
and 
\begin{equation}
\frac{I_{C}^{\left( x\right) }}{\Delta \mu }=\frac{G}{2}-\frac{2P^{2}}{%
G+\left| G^{\uparrow \downarrow }\right| ^{2}/%
%TCIMACRO{\func{Re}}%
%BeginExpansion
\mathop{\rm Re}%
%EndExpansion
G^{\uparrow \downarrow }}\stackrel{HMF}{\rightarrow }\frac{G}{4}
\end{equation}
with 
\begin{equation}
\Delta \mu _{3}^{\left( x\right) }=\frac{\Delta \mu }{2}.
\end{equation}
For a source-drain current-based HMF device very large voltage gains could
be realized, since high voltages are necessary in order to maintain a
source-drain current in the {\it off} state. The transconductance is only
one-half of the source-drain conductance, however: 
\begin{equation}
g=\left| \frac{I_{C}^{\left( x\right) }-I_{C}^{\left( z\right) }}{\Delta \mu
_{3}^{\left( x\right) }-\Delta \mu _{3}^{\left( z\right) }}\right| \stackrel{%
HMF}{\rightarrow }\frac{G}{2}
\end{equation}
It is doubtful whether such a device will be of practical use, since we have
not addressed the question of how to realize the controlled rotation of the
base terminal which should occur simultaneously with the adjustment of the
base voltage. The spin-flip transistor action therfore demonstrates nicely
the non-local action of the coherent spin-accumulation rather than being a
useful functional property.

\section{Acknowledgment}

We acknowledge discussions with Daniel Huertas-Hernando, Wolfgang Belzig, Ke
Xia as well as support by FOM and the NEDO joint research program (NTDP-98).
A.B. is supported by the Norwegian Research Council.

\end{document}